\documentclass{article}
\usepackage{amsmath}
\usepackage{graphicx}
\usepackage{slashed}

\newcommand{\bigt}{\bigtriangleup}
\usepackage{cite}
\usepackage{amssymb}
\usepackage{graphics,dcolumn,bm,float}
\usepackage{amssymb,amsmath,rotate,color}
\usepackage{booktabs}
\usepackage{rotating}
\usepackage{float}
\usepackage{braket}
\usepackage{xcolor}
\usepackage{soul}

\begin{document}
\def\boxit#1{\vcenter{\hrule\hbox{\vrule\kern8pt
      \vbox{\kern8pt#1\kern8pt}\kern8pt\vrule}\hrule}}
\def\Boxed#1{\boxit{\hbox{$\displaystyle{#1}$}}} 
\def\sqr#1#2{{\vcenter{\vbox{\hrule height.#2pt
        \hbox{\vrule width.#2pt height#1pt \kern#1pt
          \vrule width.#2pt}
        \hrule height.#2pt}}}}
\def\square{\mathchoice\sqr34\sqr34\sqr{2.1}3\sqr{1.5}3}
\def\Square{\mathchoice\sqr67\sqr67\sqr{5.1}3\sqr{1.5}3}
\def\lambdabar{{\mathchar'26\mkern-9mu\lambda}}
\def\thrdotovervx{\buildrel\textstyle...\over v_x}
\def\thrdotovervy{\buildrel\textstyle...\over v_y}
\title{\bf  Symmetries of Weyl Superconductors with Different Pairings  }

\author{{\small \ Mehran Z-Abyaneh$^1$}\footnote{mehran.z.abyaneh@gmail.com}
        {\small {},
         Mehrdad Farhoudi$^1$}\footnote{m-farhoudi@sbu.ac.ir}
        {\small {}\ and Mahdi Mashkoori$^{2,3}$}\footnote{mahdi.mashkoori@kntu.ac.ir} \\
        {\small $^1$Department of Physics, Shahid Beheshti University,
                    Evin, Tehran 19839, Iran} \\
        {\small $^2$Department of Physics, K.N.~Toosi University of Technology, P.O. Box 15875-4416, Tehran, Iran} \\
        {\small $^3$School of Physics, Institute for Research in Fundamental Sciences (IPM), P.O. Box 19395-5531, Tehran, Iran}
    }
\date{\small April 25, 2025}
\maketitle
\begin{abstract}
We examine the Bogoliubov-de Gennes Hamiltonian and its symmetries
for a time-reversal symmetry broken three dimensional Weyl
superconductor. In the limit of vanishing pairing potential, we
specify that this Hamiltonian is invariant under two sets of
continues symmetries, i.e. the $U(1)$ gauge symmetry and the
$U(1)_A$ axial symmetry. Although a pairing of the
Bardeen-Cooper-Schrieffer type spontaneously breaks both of these
symmetries, we show that a Fulde-Ferrell-Larkin-Ovchinnikov type
pairing spontaneously breaks only the $U(1)$ gauge symmetry (that
is then restored via the well-known scalar phase mode of
superconductivity). Consequently, in the former case, two
Nambu-Goldstone modes are required in the system to restore the
broken symmetries. We indicate that one of these two modes is an
emergent pseudo-scalar phase mode. We also demonstrate that such a
phase mode leads to a pseudo-Meissner effect.
\end{abstract}

\section*{Introduction}
A few decades ago, Nambu~\cite{Nambu1, Nambu2,Nambu3} suggested
that, in the Bardeen-Cooper-Schrieffer (BCS) formalism of
superconductivity~\cite{BCS}, a massless scalar collective mode
(later called the Nambu-Goldstone (NG) mode) should appear to
recover the charge conservation. That is because, although the
basic Hamiltonian of the theory, which considers only the
interactions of the electric charges, is invariant under a {\it
local} continuous $U(1)$ gauge symmetry, the mean-field reduced
BCS Hamiltonian is~not. This is an example of the spontaneous
symmetry breaking (SSB) and dynamical gap generating that reflects
the non-conservation of the electric charge. In other words, each
of the quasi particles introduced by Bogoliubov~\cite{Bogoliubov}
and Valatin~\cite{Valatin}, which are the building blocks of the
Cooper pairs, does~not appear to have a definite charge. Indeed, a
theory with broken gauge symmetry cannot describe processes
including the electromagnetic field (like the Meissner
effect~\cite{Meissner}), and as Nambu observed, the SSB of the
gauge symmetry needs to be restored by an NG mode. To restore the
symmetry, one has to take into account the radiative corrections
coming from the NG mode to the vertex diagram. After considering
this contribution, the modified vertex~\cite{Littlewood} (or the
{\it dressed} vertex) satisfies the Ward identity, and thus the
symmetry is restored via a scalar NG mode. Such a massless NG mode
(or phase mode) can also be absorbed into the longitudinal
component of electromagnetic fields, and gets elevated to the
plasma frequency due to the Anderson-Higgs
mechanism~\cite{Anderson1,Anderson2, Higgs}, which leads to the
Meissner effect.

Moreover, the similarity of the Bogoliubov-Valentin equation to
the Dirac equation~\cite{dirac2-1928} led Nambu and Jona-Lasinio
(NJL) to transform the BCS theory to strong interaction
physics~\cite{NJL1,NJL2}, wherein the global
 $U(1)_A$ symmetry [The
subscript $A$ stands for axial symmetry.] as an approximately
conserved global symmetry in flavor space is spontaneously broken.
Accordingly, the nucleon mass is generated by a SSB of the
$U(1)_A$ symmetry, and the produced pion is the pseudo-scalar NG
boson of this symmetry breaking~\cite{Abyaneh3,Abyaneh1}.

On the other hand, the correspondence between high-energy and
condensed matter physics has recently reached new levels in the
realm of novel quantum materials with the introduction of concepts
such as the Dirac and Weyl materials~\cite{Balatsky,Armitage},
which are commonly used to describe elementary
particles~\cite{Weyl(1929)}. The Weyl semimetals (WSMs) were first
proposed in the pyrochlore irridates~\cite{Wan} and later in the
heterostructures of topological and normal
insulators~\cite{Burkov}. These objects are very peculiar in the
sense that their valences and conduction bands have non-degenerate
touching points in the Brillouin zone (called Weyl nodes) whose
low-energy excitations obey the Weyl equation and are chiral
fermions. It is well-known that the Weyl nodes come in pairs of
opposite chirality~\cite{Nielsen} and are protected via
time-reversal (TR) and/or inversion (IR) symmetry~\cite{Sinha2020}
while are separated by a constant vector in momentum space.
Chirality is thus a defining emergent property of electrons in
WSMs.

Furthermore, the experimental observations and theoretical
analysis suggest a superconducting phase in WSMs, like
MoTe2~\cite{Yanpeng,Jiang2017} and TaP~\cite{Xu2015}. In addition
to intrinsic superconductivity, WSMs can also become
superconductors via the proximity effect, which occurs when one
brings a WSM close to a superconductor. It is worth mentioning
that despite the `chirality blockade' in a magnetic WSM, where the
Andreev reflection between the normal state of the WSM and a
superconductor is suppressed, introducing a Zeeman field at the
interface provides the necessary chirality switch. Indeed, such
action overcomes the blockade and enables the activation of the
Andreev reflection, leading to the proximity induced
superconductivity in WSMs~\cite{Bovenzi2017}. Also, the surface
states in WSMs, known as Fermi arc states, violate the chirality
blockade. The reason lies in the fact that Fermi arcs are~not part
of the Weyl spectrum~\cite{Faraei2019}. Moreover, it has been
reported that with a suitable choice of the Fermi cutoff, the
Fermi arcs do~not affect the paring amplitude in
WSMs~\cite{Dutta2020}. Furthermore, the classification of the
induced pairing and the magnitude of each pair amplitude has been
evaluated in the chirality blockade regime for both even- and
odd-frequency pairing~\cite{Dutta2020}. Considering the
even-frequency spin-singlet pairing, the amplitude of the
interorbital $s$-wave pairing has been shown to be $2$-orders of
magnitude larger than the intraorbital $s$-wave.

Accordingly, the Weyl equation is affected due to the presence of
superconducting term in the corresponding Hamiltonian. In this
regard, for instance, when a conventional BCS superconductor is
placed next to a WSM, an $s$-wave superconductivity is induced in
that WSM. The resulting Hamiltonian, namely the Bogoliubov-de
Gennes (BdG) Hamiltonian~\cite{Gennes,Bogoliubov1}, includes a
pairing term originated from the SSB. However, the nature of such
a pairing is under debate~\cite{Madsen2017}, as there are
generally two candidates to explain this situation. One case is
the $s$-wave Fulde-Ferrell-Larkin-Ovchinnikov (FFLO) or intranode
pairing, in which electrons are paired up on the same side of the
Fermi surface. Another is the conventional BCS or internode
pairing, in which the pair consists of electrons on opposite sides
of the Fermi surface, and the center of mass of Cooper pairs has
zero momentum. Despite some efforts, there are still debates about
the preferred pairing state in WSMs. For instance, Cho et al.
argued that the FFLO state in the IR-symmetric WSMs has
lower-energy compared to the conventional superconducting state
via the mean-field calculations~\cite{Cho}. Whereas, other groups
have demonstrated that the energy of the BCS state is lower
compared to the FFLO state~\cite{Wei,Bednik}. Both types of
pairing are allowed when the TR-symmetry is broken but the
IR-symmetry is preserved, whilst the FFLO state is the only
allowed pairing term when both of the IR-symmetry and TR-symmetry
are broken~\cite{Sinha2020}. It is worthwhile to mention that, a
minimal model of TR broken WSMs consists of a single pair of the
Weyl nodes, while a WSM with broken IR contains four Weyl nodes
with total zero chirality~\cite{Sinha2021}. It has already been
observed that in a WSM, the chirality can be understood as a
topologically protected
charge~\cite{Fang,Zhang,Jia2016,Bednik2016}, and in a
3-dimensional Weyl superconductor (3DWS), the chiral symmetry
breaking occurs in addition to gauge symmetry
breaking~\cite{Matsuda}. Indeed, in Ref.~\cite{Bednik2016}, the
authors have considered the anomalous Hall effect in a topological
Weyl superconductor with the broken TR-symmetry and demonstrated
the existence of a conserved chiral charge in WSMs. Also,
topological superconductivity in WSMs can lead to anomalies in the
presence of chiral vortex lines~\cite{Qi2013}.

In this work , we aim to provide a physical insight on the nature
of superconductivity in WSMs by focusing on symmetry
considerations. Thus, we promote the idea that in a 3DWS, besides
the $U(1)$ gauge symmetry, an emergent low-energy  $U(1)_A$
symmetry exists, which leads to a new charge for the system,
namely a chiral charge. To make this idea clearer, we study the
BdG Hamiltonian using a doubled representation of Dirac matrices.
Next, we investigate the model when the BCS- and/or the FFLO-type
pairings are proximity induced in WSMs. Then, we demonstrate that
an emergent pseudo-scalar phase mode appears in WSMs with BCS-type
superconductivity, while the FFLO-type lakes this phase mode. In
addition, we demonstrate that such a phase mode leads to a
pseudo-Meissner effect. The outline of this work is as follows. In
Sec.~2, we briefly review the SSB in the NJL model, and explain
how an NG mode appears to recover a continuous global symmetry. In
Sec.~3, we introduce the BdG Hamiltonian for a 3DWS and clearly
show that the continuous gauge and axial symmetries are
spontaneously broken via the induced pairings. Accordingly, we
indicate that, as a result of the SSB of the $U(1)_A$, an emergent
pseudo-scalar phase NG mode appears. In Sec.~4, we review the
Higgs mechanism and argue that the interaction of this
pseudo-scalar phase mode with an external pseudo-magnetic can lead
to a pseudo-Meissner effect. Finally, we furnish the conclusion in
the last section.

\section*{Chiral Invariance and New Pseudo-Scalar NG Mode }\label{sec2}
In this section, we review how the the spontaneously broken chiral
symmetry in Dirac Lagrangian can be restored by introducing a
pseudo-scalar NG mode. The Dirac Lagrangian for a free electron
with mass $m$ and momentum $\mbox{\boldmath$p$}$ is
\begin{eqnarray}\label{DiracL}
{\cal L}=\bar{\psi}(i\gamma_{\mu}\partial^{\mu}
-m){\psi},
\end{eqnarray}
where $\bar{\psi}=\psi^{\dagger}\gamma_0$, the natural units with
$\hbar=1=c$ is assumed [However, in condensed matter systems, the
Dirac Hamiltonian contains the Fermi velocity $v_{_{\rm F}}$
instead of the speed of light.] and the $\gamma_{\mu}$'s are the
Dirac gamma matrices that, in the Weyl or chiral representation,
are defined as
\begin{eqnarray}\label{D}
\gamma_{i}=\left(
                 \begin{array}{cc}
                   \mbox{\boldmath$0$} & \sigma_{i} \\
                 - \sigma_{i} & \mbox{\boldmath$0$} \\
                 \end{array}
               \right)\
               \qquad\ \textrm{and}\qquad\ \gamma_0=\left(
                 \begin{array}{cc}
                     \mbox{\boldmath$0$}& \mathbb{I}_{2\times 2}\\
                   \mathbb{I}_{2\times 2} & \mbox{\boldmath$0$} \\
                 \end{array}
               \right).
\end{eqnarray}
Here, $\sigma_{i}$ (where i=1, 2, 3) and $\mathbb{I}_{2\times 2}$
respectively are the Pauli matrices and the unit matrix, and the
symbol $\mbox{\boldmath$0$}$ represents a $2\times 2$ matrix. The
$\gamma_5$ matrix, that is constructed as
$\gamma_5=i\gamma_0\gamma_1\gamma_2\gamma_3$, is
\begin{eqnarray}\label{Dg}
 \gamma_5=\left(
                 \begin{array}{cc}
                     - \mathbb{I}_{2\times 2}&\mbox{\boldmath$0$}\\
                   \mbox{\boldmath$0$} &  \mathbb{I}_{2\times 2} \\
                 \end{array}
               \right).
\end{eqnarray}
Also, one can construct the charge-conjugation operator
$\mathcal{C}=i \gamma_2 K$, with $K$ being the complex-conjugate
operator, as
\begin{equation}
\mathcal{C}=\left(\begin{array}{cccc}
\mbox{\boldmath$0$} & i \sigma_2   \\
- i \sigma_2 & \mbox{\boldmath$0$} \\
\end{array}\right)K,
\label{chargec}
\end{equation}
which transforms a particle to an anti-particle, namely
$\mathcal{C}\psi\to\psi^{\rm c}$. Moreover, the $\gamma_{\mu}$ matrices
satisfy
\begin{equation}\{\gamma_{\mu},\gamma_{\nu}\}\equiv \gamma_{\mu} \gamma_{\nu}+\gamma_{\nu} \gamma_{\mu}=2\,\eta_{\mu\nu}
 \mathbb{I}_{4\times 4},
\label{Eq:Gamma_1}
\end{equation}
where $\eta_{\mu\nu}$ (with $\mu,\nu= 0,\cdots,3$) is the
Minkowski metric in $(1+3)$ dimensions with the signature $-2$.

The equation of motion of Lagrangian~(\ref{DiracL}) is the
celebrated Dirac equation~\cite{dirac2-1928}
\begin{equation}
i\,  \partial_{ t}\psi= \gamma_0({\mbox{\boldmath$\gamma$}}\cdot
\mbox{\boldmath$p$}+ m )\psi .
\end{equation}
By defining the left-handed and right-handed projection operators
$\psi_{\rm R}=(1+\gamma_5)\psi/2 $ and $\psi_{\rm
L}=(1-\gamma_5)\psi/2 $, the Dirac equation becomes
\begin{equation}\label{bcsdirac}
  \mbox{\boldmath$\sigma$} \cdot \mbox{\boldmath$p$} \,\psi_{\rm R} +  m \,\psi_{\rm L}= \varepsilon \psi_{\rm R}
          \qquad\ {\rm and}\qquad\
  -\mbox{\boldmath$\sigma$} \cdot \mbox{\boldmath$p$} \,\psi_{\rm L}+ m\,\psi_{\rm R}= \varepsilon \psi_{\rm L},
\end{equation}
with the  eigenvalues $ \varepsilon=\pm\sqrt{p^2+m^2 }$. Obviously
when $m=0$, the right and left chirality sectors are decoupled and
the Dirac equation reduces to the Weyl equation~\cite{Weyl(1929)}.
In this limit, Lagrangian~(\ref{DiracL}) is invariant against the
set of two independent continuous global transformations
\begin{eqnarray}\label{LNJL1}
\psi\!\!\!&\rightarrow& \!\!\!\exp{(i \alpha)}\psi\quad\qquad {\rm
hence\!:}\qquad
 \bar{\psi}\rightarrow\bar{\psi} \exp{(-i \alpha)},\\
 &&\cr
\psi\!\!\!&\rightarrow& \!\!\!\exp{(i \gamma_5\beta)}\psi\qquad
{\rm hence\!:}\qquad\bar{\psi}\rightarrow\bar{\psi}
\exp{(i\gamma_5 \beta)},
\end{eqnarray}
with $\alpha$ and $\beta$ as arbitrary constants. Due to the
Noether theorem, there are two conserved currents, namely
\begin{equation}\label{current0}
j_{\mu }  = \bar{\psi}\gamma_\mu  \psi \qquad\ {\rm and}\qquad\
 j_{5,\mu } =  \bar{\psi}\gamma_\mu\gamma_5 \psi
 \end{equation}
as the vector and axial vector currents, respectively, which
satisfy the continuity equations
\begin{equation}\label{25}
\partial^{\mu}j_{\mu}=0\qquad\ {\rm and}\qquad\ \partial^{\mu}j_{5,\mu}=0.
\end{equation}
This fact corresponds to the conservation of the electron number
and the chiral or $\gamma_5$ charge, respectively.

When the fermion mass term is generated dynamically, for example
in the context of the NJL model [This relation is referred to as
the gap equation of the NJL model~\cite{NJL1,NJL2}.], i.e. $m_{\rm
f}\propto \langle \bar{\psi}\psi\rangle$ in modified
Lagrangian~(\ref{DiracL}), the conservation relations~(\ref{25})
become
\begin{equation}\label{coservation0}
\partial^{\mu}j_{\mu }=0 \qquad\ {\rm and}\
 \qquad   \partial^{\mu}j_{5,\mu} = 2i\, m_{\rm f}
\bar{\psi}\gamma_5\psi ,
\end{equation}
which means that the mass term has spoiled the axial or the
$\gamma_5$ symmetry. Nevertheless, by including the radiative
corrections, the vertex function for the axial vector current
is~not simply given by $ \gamma_{\mu}\gamma_{5}$, instead
by~\cite{NJL1,NJL2}
\begin{eqnarray}\label{coservation5}
\lambda_{5,\mu}(\mbox{\boldmath$p$}',\mbox{\boldmath$p$})=
\gamma_{\mu}\gamma_{5}-i\frac{ 2 m_{\rm f}\gamma_5 q_{\mu}}{q^2},
\end{eqnarray}
where $\mbox{\boldmath$p$}$ and $\mbox{\boldmath$p$}'$ are the
initial and final momenta and
$\mbox{\boldmath$q$}=\mbox{\boldmath$p$}'-\mbox{\boldmath$p$}$.
Therefore, the vertex function now includes an extra term, which
indicates the existence of a pseudo-scalar zero-mass state. Hence,
by redefining $ j^{\lambda}_{5,\mu } =  \bar{\psi}\lambda_{5,\mu}
\psi$, we obtain
\begin{equation}\label{coservation02}
\partial^{\mu}j^{\lambda}_{5,\mu} = 0,
\end{equation}
which means the chiral symmetry is restored by introducing a
pseudo-scalar NG mode.

\section*{Weyl Superconductors and Chiral Symmetry}
After reviewing the concept of SSB of the $U(1)_A$ symmetry in the
context of Dirac equation, we now investigate the relevant
symmetries for WSMs after they become 3DWS as a consequence of the
SSB. In this regard, we consider a minimal relativistic low-energy
model of a WSM with two Weyl nodes of opposite chirality separated
by $2\,\mbox{\boldmath$p$} _0$ in the momentum space. The nonzero
chiral shift $\mbox{\boldmath$p$}_0$ breaks the TR-symmetry in
such a model. The Weyl Hamiltonian around the Weyl nodes at
momentum $\pm\mbox{\boldmath$p$}_0$ (in unit $v_{_{\rm F}}=1$)
reads
\begin{equation}\label{Eq.DBdG}
{\cal H}_{\rm W}(\mbox{\boldmath$p$} )=\left( \begin{array}{cc}
H^{\rm W}_{+}
& \mathbf{0} \\
\mathbf{0}  &
H^{\rm W}_{-}\\
\end{array} \right)
\qquad\ {\rm with}\qquad\ H^{\rm W}_{\pm}= \pm
\mbox{\boldmath$\sigma$}\cdot\left(\mbox{\boldmath$p$}
\mp\mbox{\boldmath$p$}_0\right),
\end{equation}
where $\mbox{\boldmath$p$} $ is the momentum of excitations and
$\pm$ stands for two chiralities of the Weyl nodes. With a proper
rotation, the term containing $\mbox{\boldmath$p$}_0 $ can be
gauged away from this Hamiltonian. In the presence of an external
electromagnetic field, such a rotation leads to an induced
$\theta$-term~\cite{ZyuzinBurkov2012} in the corresponding action
of the field, which is called the axionic field. Such a term is
responsible for the anomalous Hall effect and the chiral magnetic
effect~\cite{Burkov2}, in which the number of particles of a
specific chirality, in the presence of a topologically nontrivial
configuration of the background gauge field (like electromagnetic
field), is~not conserved. This effect is the condensed matter
counterpart of the chiral anomaly in high-energy physics. It is
also interesting to know that the chiral anomaly has been
experimentally observed in the chiral superfluid $3He-A$ with Weyl
fermionic quasiparticles\cite{Bevan}.

The pairing Hamiltonian for such a system is
\begin{equation}
{\cal H}_{\rm BdG}(\mbox{\boldmath$p$} )=\left(
 \begin{array}{cc}
{\cal H}_{\rm W}(\mbox{\boldmath$p$} )-\mu&\Delta \\
\Delta^{\dagger}&\mu-\mathcal{ C} {\cal H}_{\rm W}(\mbox{\boldmath$p$} ) \mathcal{ C}^{-1}\\
 \end{array} \right),
\label{BdGWeyl1}
\end{equation}
with $\Delta$ being the superconducting order parameter and $\mu$
the electrochemical potential. However, the chemical potential in
the Hamiltonian acts as a shift in energy and adjusting its value
leads to a quantitative change in the Fermi level of the system,
and does~not affect the output results considered in this
research. Moreover, it is common to consider an undoped WSM
corresponding to zero chemical potential, see, e.g.,
Ref.~\cite{Dutta2020}. Hence, we consider $\mu =0$ in this work.
In the literature, the two chirality sectors have been dealt with
separately, see, e.g., Ref.~\cite{Sinha2020}. However, for
convenience, we prefer to mix the two chiralities of the Weyl
nodes to construct the following Hamiltonians for the internode
and intranode pairings as
\begin{equation}
\mathcal{H}_{_{\rm BCS}}=\left(
\begin{array}{cccc}
 \mbox{\boldmath$\sigma$}\cdot\mbox{\boldmath$p$} & \mbox{\boldmath$0$}&\hat{ \bigt}_{_{\rm B}}&  \mbox{\boldmath$0$}\\
 \mbox{\boldmath$0$}&  -\mbox{\boldmath$\sigma$}\cdot\mbox{\boldmath$p$} &  \mbox{\boldmath$0$}&\hat{ \bigt}_{_{\rm B}}\\
\hat{ \bigt}^{\dagger}_{_{\rm B}} & \mbox{\boldmath$0$} & -\mbox{\boldmath$\sigma$}\cdot\mbox{\boldmath$p$} &  \mbox{\boldmath$0$}\\
 \mbox{\boldmath$0$}&\hat{ \bigt}^{\dagger}_{_{\rm B}} & \mbox{\boldmath$0$}& \mbox{\boldmath$\sigma$}\cdot\mbox{\boldmath$p$}\\
\end{array}\right)\qquad{\rm and}\qquad
\mathcal{H}_{_{\rm FFLO}}=\left(
\begin{array}{cccc}
 \mbox{\boldmath$\sigma$}\cdot\mbox{\boldmath$p$} & \mbox{\boldmath$0$}&  \mbox{\boldmath$0$}& \hat{ \bigt}_{_{\rm F}}\\
 \mbox{\boldmath$0$}&  -\mbox{\boldmath$\sigma$}\cdot\mbox{\boldmath$p$} & \hat{ \bigt}_{_{\rm F}}&  \mbox{\boldmath$0$} \\
 \mbox{\boldmath$0$} &\hat{ \bigt}^{\dagger}_{_{\rm F}} & -\mbox{\boldmath$\sigma$}\cdot\mbox{\boldmath$p$} &  \mbox{\boldmath$0$}\\
\hat{ \bigt}^{\dagger}_{_{\rm F}} & \mbox{\boldmath$0$} & \mbox{\boldmath$0$}& \mbox{\boldmath$\sigma$}\cdot\mbox{\boldmath$p$}\\
\end{array}\right),
\label{FFLO/BCS}
\end{equation}
where
\begin{eqnarray}
  \hat{\bigt}_{_{\rm B}}\! \equiv\!  \left(
\begin{array}{cccc}
\bigt_{_{\rm B}}&0\\
0& \bigt_{_{\rm B}} \\
\end{array}
\right)\qquad\ \textit{{\rm and}}\qquad\  \hat{\bigt}_{_{\rm F}}\!
\equiv\! \left(
\begin{array}{cccc}
\bigt_{_{\rm F}}&0\\
 0&\bigt_{_{\rm F}}\\
\end{array}
\right),
\end{eqnarray}
with scalars $\bigt_{_{\rm B}}$ and $\bigt_{_{\rm F}}$ that
represent the $s$-wave BCS-like and the FFLO-like pairings,
respectively. These Hamiltonians act in the space of the Nambu
spinors
\begin{eqnarray}\label{chargeeigen}
  \Phi= \left(
\begin{array}{c}
\Psi\\
 \Psi^{\rm c} \\
\end{array}
\right),
\end{eqnarray}
where $\Phi$ is a generic solution,
\begin{equation}
\Psi=\left(\psi_{-}^{\uparrow},\psi_{-}^{\downarrow},\psi_{+}^{\uparrow},\psi_{+}^{\downarrow}\right)^{\rm
T},
\end{equation}
and the charge conjugation operator is defined in~(\ref{chargec}).

As stated, Hamiltonians~(\ref{FFLO/BCS}) have the advantage of
accommodating both chiralities in a doubled representation.
However, we intend to write these Hamiltonians in a covariant
manner. Hence, for this purpose, we define the set of matrices
[Similar set of matrices has also been defined in 
Refs.~\cite{Dvoeglazov,Sokolik}.]
\begin{eqnarray}\label{newmatrices}
\Gamma_{0}  \equiv\!  \left(
\begin{array}{cccc}
  - \gamma_{0}&  \mbox{\boldmath$0$}_{4\times 4}\\
  \mbox{\boldmath$0$}_{4\times 4}& - \gamma_{0}   \\
\end{array}
\right),\
 \Gamma_{i}  \equiv\! \left(
\begin{array}{cccc}
  \gamma_{i}&  \mbox{\boldmath$0$}_{4\times 4}\\
  \mbox{\boldmath$0$}_{4\times 4}&-\gamma_{i}   \\
\end{array}
\right),\
 \Gamma_5 \equiv \!\left(
\begin{array}{cccc}
 \gamma_5 &   \mbox{\boldmath$0$}_{4\times 4}\\
  \mbox{\boldmath$0$}_{4\times 4}&  -\gamma_5 \\
\end{array}
\right),\
 \Gamma_6 \!\!\!\!\! &\equiv&\!\!\!\!\!  \left(
\begin{array}{cccc}
\mathbb{I}_{4\times 4}&   \mbox{\boldmath$0$}_{4\times 4}\\
   \mbox{\boldmath$0$}_{4\times 4} & - \mathbb{I}_{4\times 4}\\
\end{array}
\right),
\end{eqnarray}
where the $\Gamma_{\mu}$ set of matrices
form the basis of the Clifford algebra with
\begin{equation} \label{commut1}
\{\Gamma_{\mu},\Gamma_{\nu}\}= 2\, \eta_{\mu\nu}
\mathbb{I}_{8\times 8}.
\end{equation}
Accordingly, we obtain
$\Gamma_5=i\Gamma_0\Gamma_1\Gamma_2\Gamma_3$ and
\begin{equation} \label{commut2}
[\Gamma_{6}, \Gamma_{a}]=0\quad\textit{{\rm for:}\ $a=0,\cdots, 3,
5$},\qquad\qquad\  \{\Gamma_{5}, \Gamma_{\mu}\}=0.
\end{equation}
To deal with the pairing part of the BdG Hamiltonian, we also
define
\begin{eqnarray}
  \Gamma_7\! \equiv\!  \left(
\begin{array}{cccc}
  \mbox{\boldmath$0$}_{4\times 4}& \mathbb{I}_{4\times 4}\\
 \mathbb{I}_{4\times 4}&   \mbox{\boldmath$0$}_{4\times 4}\\
\end{array}
\right)\qquad\quad\textit{{\rm and}}\qquad\quad\Gamma_8\! \equiv\!
\left(
\begin{array}{cccc}
  \mbox{\boldmath$0$}_{4\times 4}&\gamma_0\\
\gamma_0&   \mbox{\boldmath$0$}_{4\times 4}\\
\end{array}
\right),
\end{eqnarray}
whose commutation relations with matrices~(\ref{newmatrices}) are
\begin{eqnarray}\label{gamma7}
\{\Gamma_7,\Gamma_{a}\}\!\!\!&=&\!\!\!0\quad\textit{{\rm for:}\
$a=1,2,3,5,6$},\qquad\qquad\  [\Gamma_7,\Gamma_{0,8}]\!=\!0, \cr
 &&\cr
[\Gamma_8,\Gamma_a]\!\!\!&=&\!\!\!0\quad\textit{{\rm for:}\
$a=0,\cdots,3,5$},\qquad \qquad\
 \{ \Gamma_8,\Gamma_{6} \}=0.
\end{eqnarray}
Using these matrices, Hamiltonians~(\ref{FFLO/BCS}) can be written
as
\begin{equation} \label{BdGHam1}
 {\cal H}_{_{\rm BCS}} =\Gamma_0\Gamma_{i} p^{i} -\Gamma_0\Gamma_8\Delta_{_{\rm B}}
\qquad\quad {\rm and}\qquad\quad {\cal H}_{_{\rm FFLO}}
=\Gamma_0\Gamma_{i} p^{i} -\Gamma_0\Gamma_7\Delta_{_{\rm F}}
\end{equation}
with
\begin{eqnarray}\label{delta}
\Delta_{_{\rm B}}\! \equiv\! \left(\begin{array}{cccc} \hat{\bigt}_{_{\rm B}}^{\dagger}&
\mbox{\boldmath$0$}&  \mbox{\boldmath$0$}&
\mbox{\boldmath$0$}\\\mbox{\boldmath$0$}& \hat{\bigt}_{_{\rm
B}}^{\dagger}&  \mbox{\boldmath$0$}& \mbox{\boldmath$0$}\\
\mbox{\boldmath$0$}& \mbox{\boldmath$0$}& \hat{\bigt}_{_{\rm B}}&
\mbox{\boldmath$0$}\\ \mbox{\boldmath$0$}& \mbox{\boldmath$0$}&
\mbox{\boldmath$0$}& \hat{\bigt}_{_{\rm B}}\\\end{array}\right)
\quad\qquad\textit{{\rm and}}\quad\qquad\Delta_{_{\rm F}}\!
\equiv\! \left(\begin{array}{cccc}
\hat{\bigt}_{_{\rm F}}^{\dagger}& \mbox{\boldmath$0$} &  \mbox{\boldmath$0$}& \mbox{\boldmath$0$}\\
\mbox{\boldmath$0$}&  \hat{\bigt}_{_{\rm F}}^{\dagger}&  \mbox{\boldmath$0$}& \mbox{\boldmath$0$}\\
\mbox{\boldmath$0$}&  \mbox{\boldmath$0$}& \hat{\bigt}_{_{\rm F}}& \mbox{\boldmath$0$}\\
\mbox{\boldmath$0$}&  \mbox{\boldmath$0$}& \mbox{\boldmath$0$}& \hat{\bigt}_{_{\rm F}}\\
\end{array}\right).
\end{eqnarray}
Based on Refs.~\cite{peskin,salehi}, as the FFLO-like superconducting pairing is given by
$\bar\psi\Delta_{_{\rm F}}\gamma^0\psi_c$ and the
BCS-like pairing is given by $\bar\psi\Delta_{_{\rm
B}}\mathbb{I}_{4\times 4} \psi_c$, hence the Lorentz invariance is
satisfied.

Now, the Lagrangian of the system can be written as
\begin{equation} \label{BdGlag1}
{\cal L}_{_{\rm BCS }} =\bar{\Phi}(\Gamma_{\mu} p^{\mu}
-\Gamma_8\Delta_{_{\rm B}})\Phi\qquad\qquad {\rm and}\qquad\qquad
 {\cal L}_{_{\rm FFLO }}=\bar{\Phi}(\Gamma_{\mu} p^{\mu} -\Gamma_7\Delta_{_{\rm
 F}})\Phi,
\end{equation}
where $\bar{\Phi}\equiv\Phi^{\dagger}\Gamma_0$. Hence, the
corresponding equations of motions are
\begin{equation}\label{BdGeq1}
 (\Gamma_{\mu} p^{\mu} -\Gamma_8\Delta_{_{\rm
B}})\Phi =0\qquad\qquad {\rm and/or}\qquad\qquad
 (\Gamma_{\mu} p^{\mu} -\Gamma_7\Delta_{_{\rm F}})\Phi=0.
\end{equation}

To get a better sense on the role of $\Gamma_5$ and $\Gamma_6$, we
also define the projection operators
\begin{equation}\label{projection}
P_{\pm}^5=\frac{1}{2}(\mathbb{I}\pm\Gamma_5)\qquad\quad{\rm
and}\qquad\quad P_{\pm}^{6}=\frac{1}{2}(\mathbb{I}{\pm}\Gamma_6),
\end{equation}
where $ \mathbb{I}$ is $8\times 8$ identity matrix, and we have
$P_{\pm}^{5^{2}}=P_{\pm}^5,\ P_{\pm}^{6^{2}}=P_{\pm}^6$. In addition,
acting these operators on the Nambu spinors~(\ref{chargeeigen})
yields
\begin{eqnarray}\label{23}
P_{+}^5 P_{+}^6 \Phi
\!\!\!\!&=&\!\!\!\!\frac{1}{4}(\mathbb{I}+\Gamma_5)(\mathbb{I}+\Gamma_6)\Phi=\Psi_+
, \qquad\quad
P_{+}^5 P_{-}^6 \Phi= \frac{1}{4}(\mathbb{I}+\Gamma_5)(\mathbb{I}-\Gamma_6)\Phi=\Psi^c_+ , \qquad \\
P_{-}^5 P_{+}^6\Phi
\!\!\!\!&=&\!\!\!\!\frac{1}{4}(\mathbb{I}-\Gamma_5)(\mathbb{I}+\Gamma_6)\Phi=\Psi_-
, \qquad\quad P_{-}^5 P_{-}^6 \Phi
=\frac{1}{4}(\mathbb{I}-\Gamma_5)(\mathbb{I}-\Gamma_6)\Phi=\Psi^c_-
.
\end{eqnarray}
These relations suggest that the operator $\Gamma_5$ is related to
the chirality and $\Gamma_6$ is related to the particle-hole
symmetry.

In the limit $\Delta_{_{\rm B}}=0=\Delta_{_{\rm F}}$,
Hamiltonians~(\ref{BdGHam1}) will obviously be the same and will
be invariant against the set of two independent continuous global
transformations
\begin{eqnarray}  \label{gaugetrans}
&&\Phi({\bf r}, t) \to e^{i\,\Gamma_5 \theta/2}\Phi({\bf r},
t)\qquad {\rm hence\!:}\qquad \bar{\Phi}({\bf r}, t) \to
\bar{\Phi}({\bf r}, t) e^{i\,\Gamma_5\theta/2},\cr
 &&\cr
 && \Phi({\bf r}, t)
\to e^{i\,\Gamma_6\varphi/2} \Phi({\bf r}, t)\qquad {\rm
hence\!:}\qquad \bar{\Phi}({\bf r}, t) \to  \bar{\Phi}({\bf r}, t)
e^{-i\,\Gamma_6 \varphi/2}
\end{eqnarray}
that can also be written as
 \begin{align}
\label{chargeeigentrans} &\Phi \to \left(\Psi_-  e^{-i\,
\theta/2},\Psi_+ e^{i\,\theta/2},\Psi_+^c e^{i\,\theta/2},\Psi_-^c
e^{-i\,\theta/2}\right)^{\rm T},\cr
 &\Phi \to \left(\Psi_-
e^{i\,\varphi/2},\Psi_+ e^{i\,\varphi/2},\Psi_+^c
e^{-i\,\varphi/2},\Psi_-^c e^{-i\,\varphi/2}\right)^{\rm T},
\end{align}
where $\theta$ and $\varphi$ are arbitrary constants.
The Noether theorem dictates that such invariance leads to the
conserved currents
\begin{equation}
J_{5,\mu }  = \bar{\Phi}\Gamma_{\mu} \Gamma_5\Phi \qquad\quad{\rm
and}\qquad\quad J_{6,\mu } = \bar{\Phi}\Gamma_{\mu}
\Gamma_{6}\Phi.
\end{equation}
These currents satisfy the continuity equations
\begin{equation}\label{conserv25}
\partial^{\mu}J_{5,\mu}=0\,\qquad\quad{\rm and}\qquad\quad\partial^{\mu}J_{6,\mu}=0 .
\end{equation}
The $J_{5,\mu}$ (which originates from $\Gamma_5$ that is related
to the axial symmetry) is the axial current, and in the same vein,
the $J_{6,\mu }$ is the electromagnetic current.

It is noteworthy that the continuous global transformation leading
to $J_{5,\mu}$ also gives rise to a $\theta$ phase shift between
the pairings at Fermi surfaces with opposite chiralities. Such a
phase shift may lead to interesting observable effects in the
Josephson phenomena~\cite{Josephson1962}. Further discussion of
this topic is beyond the scope of this work and will be presented
in upcoming works. However, it should be noted that our work
differs from the Leggett work~\cite{Leggett1966,Rossi} because for
the Leggett mode to appear, it is necessary to have an interband
pairing term that couples positive and negative chiralities.
Whereas this is~not the case in Eqs.~(\ref{BdGeq1}), where we
consider either internode pairing, $\bigt_{_{\rm B}}$, or
intranode pairing, $\bigt_{_{\rm F}}$, and do~not assume both
pairings simultaneously.

Now, by utilizing transformations~(\ref{gaugetrans}) and
commutation relations (\ref{gamma7}), we observe that
$\Delta_{_{\rm B}}$ in the left-hand Hamiltonian~(\ref{BdGHam1})
breaks both of the $\Gamma_5$ and $\Gamma_6$ symmetries,
 while $\Delta_{_{\rm F}}$ in the right-hand Hamiltonian~(\ref{BdGHam1}) breaks
only the $\Gamma_6$ symmetry. Accordingly, when only
$\Delta_{_{\rm F}}$ is present in the system, we can justify that
it transforms as $\Delta_{_{\rm F}}\rightarrow  \Delta_{_{\rm
F}}\, e^{i\,\Gamma_6\varphi}$, and hence the conservation
relations~(\ref{conserv25}) become
\begin{equation}\label{coservation1}
\partial^{\mu}J_{5,\mu }=0\qquad\quad{\rm and}\qquad\quad \partial^{\mu}J_{6,\mu}=-2i\, \Delta_{_{\rm
F}}\bar{\Phi}\Gamma_7\Gamma_6\Phi.
\end{equation}
Whereas, when only $\Delta_{_{\rm B}}$ is present, it transforms
as $\Delta_{_{\rm B}}\rightarrow \Delta_{_{\rm B}}\,
e^{i\,\Gamma_6\varphi+ i\,\Gamma_5\theta}$, and one has
\begin{equation}\label{coservation2}
\partial^{\mu}J_{5,\mu }=-2i\, \Delta_{_{\rm
B}}\bar{\Phi}\Gamma_8\Gamma_5\Phi\qquad\quad {\rm
and}\qquad\quad\partial^{\mu}J_{6,\mu}=-2i\, \Delta_{_{\rm
B}}\bar{\Phi}\Gamma_8\Gamma_6\Phi.
\end{equation}
As is obvious, when $\Delta_{_{\rm B}}$ is non-zero, an emergent
pseudo-scalar phase mode should appear to restore the broken
$\Gamma_5$ symmetry.

Following the same procedure that led to
relation~(\ref{coservation5}), when only $\Delta_{_{\rm F}}$ is
present, the vertex for $J_{5,\mu }$ and $J_{6,\mu }$ currents
become
\begin{eqnarray}\label{onlyF}
\Lambda_{5,\mu}^{\rm F}
(\mbox{\boldmath$p$}',\mbox{\boldmath$p$})= \Gamma_{\mu}\Gamma_5\,
 \qquad \quad{{\rm and}}\qquad\quad
\Lambda_{6,\mu}^{\rm F}(\mbox{\boldmath$p$}',\mbox{\boldmath$p$})
= \Gamma_{\mu} \Gamma_6+i \frac{2 \Delta_{_{\rm F}}\Gamma_7 \Gamma_6
q_{\mu}}{q^2} ,
\end{eqnarray}
and when only $\Delta_{_{\rm B}}$ is the non-vanishing one, we
obtain
\begin{eqnarray}\label{onlyB}
\Lambda_{5,\mu}^{\rm B}
(\mbox{\boldmath$p$}',\mbox{\boldmath$p$})=
\Gamma_{\mu}\Gamma_5+i \frac{2 \Delta_{_{\rm B}}\Gamma_8 \Gamma_5
q_{\mu}}{q^2}\,
 \qquad \quad{{\rm and}}\qquad\quad
\Lambda_{6,\mu}^{\rm B}(\mbox{\boldmath$p$}',\mbox{\boldmath$p$})
=  \Gamma_{\mu} \Gamma_6+i \frac{2 \Delta_{_{\rm B}}\Gamma_8 \Gamma_6
q_{\mu}}{q^2} .
\end{eqnarray}
Hence for conditions~(\ref{onlyB}), new conserved currents, say
$J^{\rm B}_{5,\mu}$ and  $J^{\rm B}_{6,\mu}$, satisfy the
conservation relations
\begin{equation}\label{coservation6}
\partial^{\mu}J^{\rm B}_{5,\mu}\equiv\partial^{\mu}\bar{\Phi}\Lambda_{5,\mu}^{\rm B}\Phi=0 \qquad\quad{\rm
and}\qquad\quad
\partial^{\mu}J^{\rm B}_{6,\mu}\equiv\partial^{\mu}\bar{\Phi}\Lambda^{\rm B}_{6,\mu}\Phi=0.
\end{equation}
Also, for conditions~(\ref{onlyF}), we have
\begin{equation}\label{coservation7}
\partial^{\mu}J^{\rm F}_{5,\mu}\equiv\partial^{\mu}\bar{\Phi}\Lambda^{\rm F}_{5,\mu}\Phi=0 \qquad\quad{\rm
and}\qquad\quad
\partial^{\mu}J^{\rm F}_{6,\mu}\equiv\partial^{\mu}\bar{\Phi}\Lambda^{\rm F}_{6,\mu}\Phi=0 .
\end{equation}
These conservation relations demonstrate how the NG modes restore
the broken symmetries. Thus, the FFLO-like pairing only breaks the
gauge symmetry. Therefore, only a scalar phase mode appears to
restore the conservation of this symmetry. Whereas, the BCS-like
pairing breaks both of the gauge and the chiral symmetries, and
leads to an extra second NG mode.

To be able to detect this new emergent pseudo-scalar NG mode, we
recall that a scalar phase mode can interact with an external
photon field, leading to the Meissner effect due to the
Anderson-Higgs mechanism~\cite{Anderson1,Anderson2, Higgs}.
Actually, as Weinberg stated ``a superconductor is simply a
material in which electromagnetic gauge invariance is
spontaneously broken~\cite{weinburgssb}". In the same vein, we
expect a phenomenon similar to the Meissner effect, in which the
pseudo-scalar NG mode (created by the SSB of the chiral symmetry
via $\Delta_{_{\rm B}}$) gets absorbed by an external
pseudo-magnetic field, leading to the field expulsion, and an
effect we refer to as pseudo-Meissner.

Such a pseudo-magnetic field can be realized in WSMs in the form
of elastic gauge fields constructed with the deformation tensor by
coupling the lattice deformations to electronic degrees of
freedom~\cite{Ilan}. The most interesting feature of these elastic
gauge fields is that they are axial pseudo-gauge fields and couple
to opposite chiralities with opposite signs. The presence of
elastic gauge fields in WSMs was first predicted in
Ref.~\cite{Cortijo:2016yph} and recently realized
experimentally~\cite{KS19}. Further works on the analysis of their
physical consequences can be found in
Refs.~\cite{PCF16,GVetal16,Cortijo:2016wnf,CZ16,LPF17,GMetal17b,GMetal17a,ACV17,GMetal17c,GMSS17,AV18}.
In general, such an axial gauge field can be described
as~\cite{Cortijo:2016yph,Chernodub}
\begin{eqnarray}
A_{\mu}^5 = \xi v_{\mu\nu}\, b^{\nu},
\end{eqnarray}
where
\begin{eqnarray}
v_{\mu\nu}(x) = \frac{1}{2}(\partial_{\mu}
v_{\nu}+\partial_{\nu}v_{\mu}),
\end{eqnarray}
is the strain tensor, $v^{\mu}$ is the displacement vector, $\xi$
is a Gruneisen parameter and the vector $b^{\mu}$ quantifies the
separation of Weyl nodes in momentum space. The pseudo-vector
field ${\bf{A}}^5$ can be attributed to an effective axial vector
potential~\cite{Sukhachov1,Sukhachov2} . Then the axial magnetic
and electric fields can be defined as
$B^{5,\mu}=\frac{1}{2}\epsilon^{\mu\nu\gamma}\,\partial_{nu}A_{\gamma}^5$
and $E_{\mu}^{5}=-\partial_t A_{mu}^5$, respectively. Apparently,
the physical properties of the lattice deformation, such as
ripples or strains, determine the size of the emergent
pseudo-field in this system~\cite{Ilan}. In principle, one can
tune the pseudo-field by controlling the lattice deformation,
which was previously studied in Dirac materials
theoretically~\cite{Amorim} and observed
experimentally~\cite{Levy}.

The pseudo-magnetic (or an axial magnetic) field $\mathbf{B}^5$ is
an observable quantity, whose magnitude typically varies from
about $0.3~\mbox{T}$ to about $15~\mbox{T}$. It couples to
fermions of opposite chirality with different sign and is
therefore a chiral field. It also gives rise to an unusual
dynamics of Cooper pairs, where no Meissner effect is present,
see, e.g., Ref.~\cite{Fujimoto}. In fact, this field is completely
different from the usual electromagnetic field and does~not induce
diamagnetic currents (which can destroy
superconductivity)~\cite{Sukhachov1,Sukhachov2}. Now, taking into
account the possibility that the pseudo-scalar phase mode can be
absorbed by the ${\bf{B}}^5$ field, the pseudo-Meissner effect is
a plausible consequence of the local chiral symmetry breaking in
3DWS. We investigate this case in the next section.

As another peculiarity of the emergent pseudo-scalar phase mode,
we infer that this mode may interact with the axial gauge bosons
of the standard model of particle
physics~\cite{Glashow,Salam,Weinberg}, which also leads to the
pseudo-Meissner effect.

\section*{Pseudo-Meissner Effect}
In this section, via the Higgs mechanism, we study the process in
which the pseudo-scalar NG mode gets absorbed by the
pseudo-magnetic field, and results in the field expulsion. In this
regard, we first introduce the Abelian Higgs mechanism, whose
Lagrangian is
\begin{equation}\label{abelian1}
{\cal{L}}_\eta =
 -\frac{1}{4}F_{\mu\nu}F^{\mu\nu} +
 |(i\partial_\mu -  e A_\mu)\eta|^2 - V(\eta),
\end{equation}
where the field $\eta$ is a complex classical field over
spacetime, $e$ is the unit of the electric charge,  $F^{\mu\nu}
=\partial^{\mu} A^{\nu} -\partial^{\nu} A^{\mu} $ and $A^{\mu} $
is the electromagnetic vector potential. Meanwhile, the potential
energy for the field $\eta$ is
\begin{equation}
\label{pot1}
V(\eta) = - M^2|\eta|^2  + \frac{1}{2} \lambda |\eta|^4,
\end{equation}
where $M$ and $\lambda$ are constants and the Lagrangian
${\cal{L}}_\eta $ remains invariant under the Abelian $U(1)$ gauge
transformation $A_\mu \rightarrow A_\mu -
\partial_\mu\chi(x)/e$ with $ \eta \rightarrow e^{i\,\chi(x)}
\eta$, wherein $\chi(x)$ is a real field without dimension in the
natural units.

At this stage, we arrange to have a `spontaneous breaking of the
gauge symmetry' by choosing $<{\eta}> = v/\sqrt{2} $, where $v$ is
the vacuum expectation value of the field $\eta$ and we set it to
be real without loss of generality.  The potential energy is
minimized when
\begin{equation}
v = \frac{M}{\sqrt{\lambda}}.
\end{equation}
Then, we fix the gauge such that
\begin{equation}\label{manifest1}
\eta  =  {\frac{v e^{i\,\chi(x)}}{\sqrt 2}},
\end{equation}
where we have ignored an excitation of the potential around its
minimum for the sake of convenience. Accordingly, in order to
fully fix the gauge, we introduce a new vector potential field
\begin{equation}\label{manifest}
  A_\mu \rightarrow B_\mu  =  A_\mu - \frac{1}{e}\partial_\mu\chi(x),
\end{equation}
with which Lagrangian (\ref{abelian1}) leads to~\cite{Fraser}
\begin{equation}
\widetilde{\cal{L}}_\eta =
 -\frac{1}{4}B_{\mu\nu}B^{\mu\nu}
+ \frac{1}{2} e^2 v^2 B_\mu B^\mu,
\end{equation}
where  $B_{\mu\nu}=\partial_{\mu}B_{\nu}- \partial_{\nu} B_{\mu}$.
Lagrangian $\widetilde{\cal{L}}_\eta $ is interpreted as
containing a massive vector field $B_{\mu}$ with the mass $m_{\rm
ph}=e\, v$, where the subscript ${\rm ph}$ stands for photon. The
mass of the magnetic field is inversely proportional to its
penetration depth, $\lambda$, i.e.,
\begin{equation}\label{lambda1}
\frac{1}{m_{\rm ph}}=\lambda = \sqrt{\frac{m^*}{4\, \mu_0\,
e^2<\!\!\eta\!\!>^2 }},
\end{equation}
where $m^*$ is the effective mass of the Cooper pairs and $\mu_0$
is the permeability of free space.

This model is a Lorentz invariant version of the Landau-Ginzburg
(LG) model of superconductivity~\cite{Higgs}, which is a
phenomenological approach to describe the macroscopic properties
of superconductors, and it can be said that the NG boson is
`eaten' to become the longitudinal degree of freedom of the
photon. We know that in the LG model, the order parameter is a
complex scalar field proportional to the BCS pairing potential. We
also expect the order parameter to be proportional to the pairing
potentials of the model. To integrate our results with the Higgs
mechanism, we need to introduce two matrix form order parameters,
say $\eta_{\rm B}\equiv \Delta_{\rm B} $ and $\eta_{\rm F}\equiv
\Delta_{\rm F}$, where $\Delta_{\rm B}$ and $\Delta_{\rm F}$ are
given in relation~(\ref{delta}) and hence, $\eta_{\rm F}$ and
$\eta_{\rm B}$ transform as $\eta_{\rm F}\rightarrow \eta_{\rm
F}\, e^{i\,\Gamma_6\varphi}$ and $\eta_{\rm B}\rightarrow
\eta_{\rm B}\, e^{i\,\Gamma_6\varphi+ i\,\Gamma_5\theta}$. Then,
besides $A_{\mu}$, we also use $A_{\mu}^5$ as a
pseudo-electromagnetic field, and assume that both of these fields
transform as
\begin{eqnarray}  \label{gauge1}
&& A_{\mu}\rightarrow A_{\mu} -\frac{1}{e}
\partial_\mu\chi\quad\, \qquad{\rm with\!:}\qquad
\begin{cases}
\eta_{\rm B}({\bf r}, t) \to \eta_{\rm B}({\bf r}, t)
e^{i\,\Gamma_6\chi}\\
\eta_{\rm F}({\bf r}, t) \to \eta_{\rm F}({\bf r}, t)
e^{i\,\Gamma_6\chi},
\end{cases}
 \cr
 &&\cr
&& \! A_{\mu}^5 \rightarrow A_{\mu}^5 -\frac{1}{g}
\partial_\mu\chi\qquad{\rm with\!:}\qquad
\begin{cases}
{\eta}_{\rm B}({\bf r}, t) \to {\eta}_{\rm B}({\bf r}, t)
e^{i\,\Gamma_5\chi}\\
 {\eta}_{\rm F}({\bf r}, t) \to {\eta}_{\rm F}({\bf r}, t),
\end{cases}
\end{eqnarray}
where $g$ is a pseudo-charge. Therefore, the related invariant
Higgs Lagrangians under these transformations are
\begin{equation}\label{abelian2}
{\cal{L}}_{\rm B}  =
 -\frac{1}{4}F_{\mu\nu}F^{\mu\nu}  +
{\rm trace}\, ( |( \mathbb{I}_{8\times 8}i\partial_\mu -
e\Gamma_6 A_\mu)\eta_{\rm B}|^2 )- V(\eta_{\rm
 B}),
\end{equation}
\begin{equation}\label{abelian3}
{\cal{L}}_{\rm F}  =
 -\frac{1}{4}F_{\mu\nu}F^{\mu\nu}  +
{\rm trace}\, ( |( \mathbb{I}_{8\times 8}i\partial_\mu -  e
\Gamma_6 A_{{\mu}})\eta_{\rm F}|^2) - V(\eta_{\rm F})
\end{equation}
and
\begin{equation}
 {\cal{L}}_{5,{\rm B}}  =
  -\frac{1}{4}F_{5,{\mu\nu}}F_5^{\mu\nu} +
{\rm trace}\, ( |(\mathbb{I}_{8\times 8}i\partial_\mu -   g
\Gamma_5 A_{5,{\mu}})\eta_{\rm B}|^2) - V(\eta_{\rm
 B}),
\end{equation}
where $F_5^{{\mu\nu}} =\partial^{\mu} A_5^{\nu}  -\partial^{\nu}
A_5^{\mu} $. Analogously, by using the new vector
potential~(\ref{manifest}) and defining  $B_5^{{\mu\nu}}\!\!
=\!\partial^{\mu} B_5^{\nu}  -\partial^{\nu} B_5^{\mu} $, these
Lagrangians lead to
\begin{equation}
\widetilde{\cal{L}}_{\eta_{\rm B}}  =
 -\frac{1}{4}B_{\mu\nu}B^{\mu\nu}
+ \frac{1}{2} e^2 v^2 B_\mu B^\mu, \qquad\qquad\qquad
\widetilde{\cal{L}}_{\eta_{\rm F}}  =
 -\frac{1}{4}B_{\mu\nu}B^{\mu\nu}
+ \frac{1}{2} e^2 v^2 B_\mu B^\mu
\end{equation}
and
\begin{equation}
 \widetilde{\cal{L}}_{5,{\eta_{\rm
B}}}  =
 -\frac{1}{4}B_{5,{\mu\nu}}B_5^{\mu\nu}
+ \frac{1}{2} g^2 v^2 B_{5,{\mu}} B_5^{\mu},
\end{equation}
which indicate that the BCS type of pairing causes the
pseudo-magnetic field to acquire the mass $m^p_{\rm ph}=g\, v$,
where the superscript $p$ stands for pseudo-magnetic field. In
other words, it leads to a pseudo-Meissner effect, which hopefully
provides an achievable and quantitatively testable phenomenon to
detect the emergent pseudo-scalar NG mode. Similar to the Meissner
effect, the mass of the pseudo-magnetic field is inversely
proportional to its penetration depth, $\lambda^{p}$, i.e.,
\begin{equation}
\frac{1}{m^p_{\rm ph}}=\lambda^{p} = \sqrt{\frac{m^*}{4\,
\mu_{0}^p\, g^2 <\!\!\eta_{\rm B}\!\!>^2}},
\end{equation}
where $ \mu_{0}^p$ is the permeability of free space for the
pseudo-magnetic field. Assuming that the coupling $g$ and
$\mu_{0}^p$ are in the same order as those of the electromagnetic
interactions~\cite{Ilan}, the order of $\lambda^{p}$ will be the
same as the normal Meissner effect, i.e., $\lambda$ in
relation~(\ref{lambda1}) in topological
superconductors~\cite{Radmanesh}.

\section*{Summary and Conclusions}
 \setcounter{equation}{0}
 \renewcommand{\theequation}{A.\arabic{equation}}
In this work, we have considered a conventional BCS superconductor
is placed next to a WSM, where consequently an $s$-wave
superconductivity is induced in that WSM due to the proximity
effect. To shed light on the nature of the induced pairing, we
have considered the continuous symmetries of a 3DWS. In the
literature, there have been discussions about whether the
superconducting state in that WSM is of the FFLO or BCS type.
Here, we have shown that unlike the orthodox BCS superconductors,
wherein the $U(1)$ gauge symmetry is the only symmetry of the
gapless Hamiltonian (which is then spontaneously broken by the
dynamically generated $s$-wave BCS-like pairing), the gapless
Hamiltonian of a 3DWS is invariant under two symmetries, namely
the $U(1)$ gauge and $U(1)_A$ axial symmetries. To better
investigate the issue, we have written the BdG Hamiltonian using
the doubled representation of Dirac matrices, which has enabled us
to introduce two generators to represent the symmetries of the
system. Consequently, two charges appear in the system, i.e. the
electric charge and the chiral charge.

Furthermore, we have demonstrated that the dynamical generation of
the $s$-wave BCS-like pairing, $\bigt_{_{\rm B}}$, breaks both the
symmetries spontaneously, whereas the FFLO-like pairing,
$\bigt_{_{\rm F}}$, breaks only the $U(1)$ gauge symmetry.
Nevertheless, we have indicated that the conservation of both
charges get restored by introducing the NG modes. That is, when
only the $U(1)$ gauge symmetry gets broken, the well-known scalar
mode recovers the charge conservation. Whereas, when both of the
$U(1)$ gauge and $U(1)_A$ axial symmetries are broken, apart from
the scalar mode, one needs to introduce an extra new pseudo-scalar
phase mode to restore the $U(1)_A$ symmetry as well.

Analogous to conventional superconductors, where the scalar phase
mode of superconductivity leads to the Meissner effect in the
presence of a normal magnetic field, we have demonstrated that a
similar effect appears in 3DWSs, due to the interaction of the
pseudo-scalar phase mode with an external pseudo-magnetic. Indeed,
the obtained results indicate that when the $U(1)_A$ symmetry is
broken by $\Delta_{_{\rm B}}$, the corresponding emergent
pseudo-scalar phase mode can get absorbed by an external
pseudo-magnetic field (i.e., a ${\bf{B}}^5$ field) that leads to
an effect analogous to the Meissner effect. We have referred to
this effect as a pseudo-Meissner effect, and it can be tested in
future experiments as a key prediction of this work. More
specifically, we expect the repulsion of the pseudo-magnetic field
inside the WSMs in the superconducting s-wave phase. The
pseudo-Meissner effect is consequently achieved by a
superconducting surface current that produces a secondary
pseudo-magnetic field to compensate for the primary
pseudo-magnetic field.
 Also, the emergent pseudo-scalar NG mode
may, in principle, interact with the axial gauge bosons of the
standard model of particle physics, and leads to the
pseudo-Meissner effect by expelling these bosons from the 3DWS.
These predictions are testable in present and forthcoming
experiments. As the pseudo-Meissner effect emerges from the
BCS-like pairing in WSMs, we propose the pseudo-Meissner effect as
a useful tool for distinguishing between the FFLO-like and the
BCS-like pairings.

A possible extension of this work is to study the effect of the
new phase mode in phenomenon such as the Josephson effect, which
we will investigate in upcoming works. Also, a Weyl superconductor
originating from a WSM with broken inversion symmetry can lead to
various physical effects such as response to external magnetic
fields, thermal conductance, and transport properties, which can
be investigated in a different work. However, a recent publication
has considered the proximity effect on an inversion broken
WSM~\cite{Dawson}.


%
\end{document}